\documentclass{article}

\usepackage{graphicx}
\usepackage{amsmath}
\usepackage{amssymb}
\usepackage{epsfig}
\usepackage{bm}
\newcommand{\ket}[1]{\ensuremath{\left|{#1}\right\rangle}}
\newcommand{\bra}[1]{\ensuremath{\left\langle{#1}\right |}}

\newcommand{\rmvect}[1]{\boldsymbol{\mathrm{#1}}}
\newcommand{\brm}[1]{\boldsymbol{\mathrm{#1}}}
\newcommand{\bsy}[1]{\boldsymbol{#1}}
\newcommand{\vect}[1]{\boldsymbol{#1}}
\newcommand{\oper}[1]{\boldsymbol{\mathsf{#1}}}
\newcommand{\sinc}{\ensuremath{\mathrm{sinc}}}

\begin{document}

\markboth{Walborn, Nogueira, de Oliveira, P\'adua and Monken}{Multimode Hong-Ou-Mandel Interferometry}

%

%

\title{Multimode Hong-Ou-Mandel Interferometry}
\author{\footnotesize S. P. Walborn\footnote{Corresponding author:  swalborn@if.ufrj.br} \footnote{Current address:  Instituto de F\' \i sica, Universidade Federal do Rio de Janeiro, Caixa Postal 68528, Rio de Janeiro, RJ 21945-970, Brazil}, W. A. T. Nogueira, A. N. de Oliveira, S. P\'adua, C. H. Monken}

\maketitle

\begin{abstract}
We review some recent experiments based upon multimode two-photon interference of photon pairs created by spontaneous parametric down-conversion.  The new element provided by these experiments is the inclusion of the transverse spatial profiles of the pump, signal and idler fields.  We discuss multimode Hong-Ou-Mandel interference, and show that the transverse profile of the pump beam can be manipulated in order to control two-photon interference.  We present the basic theory and experimental results as well as several applications to the field of quantum information.        

\end{abstract}

\section{Introduction}

\begin{figure}[th]
\centerline{\psfig{file=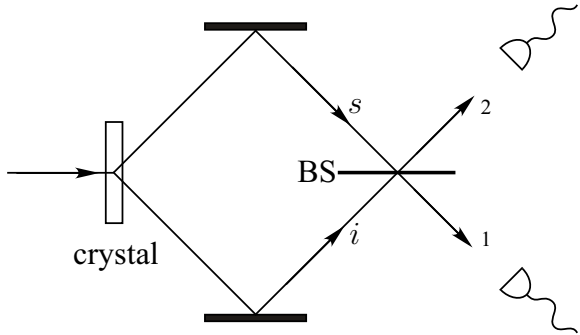,width=3in}}
\vspace*{8pt}
\caption{ \label{fig:hom_simples} Illustration of a HOM interferometer.}
\end{figure}

Fourth-order interference of two photons at a beam splitter was first studied by Hong, Ou and Mandel (HOM)\cite{hom87}, as a method to measure the time interval between photons created by spontaneous parametric down-conversion (SPDC) and consequently the coherence length of the photon wave packet to a high degree of accuracy.  It has since played a central role in a wide variety of experiments, ranging from tests of quantum mechanics versus local realism\cite{shih88,torgerson95}, to measurements of the single-photon tunneling time\cite{steinberg93}, to quantum information tasks such as Bell-state measurements \cite{braunstein95,mattle96} and two-photon controlled logic operations \cite{klm01,ralph02}, among others.    A typical HOM interferometer is shown in figure \ref{fig:hom_simples}.  Two photons are created by SPDC and sent to opposite sides of a beam splitter (BS).   In their original work, HOM showed that when the path lengths of arms $s$ and $i$ are adjusted so that the wave packets of identical photons in a symmetric polarization state overlap perfectly at a beam splitter, the photons leave the beam splitter through the same output port.  Likewise, photons in the antisymmetric singlet polarization state leave the beam splitter in different outputs.  It has been pointed out that this interference behavior can be understood by considering the overall bosonic nature of the two-photon state\cite{zeilinger94}.  
\par
Until recently, the majority of works regarding HOM interference concerned an ideal monomode (one spatial mode) situation.
Here we discuss some recent experiments involving multimode HOM interference\cite{walborn03a}.  It is well known that SPDC allows the creation of photons entangled in several degrees of freedom, including, for example, polarization and momentum (spatial mode).   Moreover, the down converted photons can be \emph{hyperentangled}:  entangled in two or more degrees of freedom simultaneously\cite{kwiat97}.  Considering fields with multiple spatial modes provides additional degrees of freedom that can be used to control and manipulate two-photon interference.  We will first review the two-photon quantum state generated by multimode SPDC.  In section \ref{sec:mmhom} we provide the basic theory of multimode HOM interference and discuss the experimental results.  We will then present some recent applications such as transverse phase measurements of the multimode two-photon field, non-classical two-photon beams and Bell-state measurements.
\section{The Two-photon Quantum State}
\label{sec:state}
Here we review the two-photon state generated by SPDC.  A more detailed account can be found elsewhere\cite{hong85,monken98a}.
For a sufficiently weak cw laser, the quantum state generated by SPDC can be shown to be
\begin{equation}
\ket{\psi}_{12} = C_{1}\ket{\mathrm{vac}} + C_{2}\ket{\psi},
\end{equation}
where $\ket{\mathrm{vac}}$ is the vacuum state and $\ket{\psi}$ is a two-photon state.  The coefficients $C_{1}$ and $C_{2}$ are such that $|C_{2}| \ll \, |C_{1}|$.  Under most experimental situations it is reasonable to work in the paraxial approximation\cite{saleh91}, in which it is assumed that the fields are appreciable only in a region that is very close to the propagation axis.  In other words, the magnitude of the transverse component of the wave vector is much smaller than the magnitude of the wave vector itself :  $|\brm{q}|  \ll \,  |\brm{k}|$.  Coordinate systems are chosen such that the down-converted fields are propagating along the $z$-direction.  It is further assumed that the experimental setup incorporates narrow bandwidth interference filters centered at twice the pump beam wavelength $\lambda_{p}$.  This allows us to consider only monochromatic down-converted fields with wavelength $\lambda_{c}=2\lambda_{p}$.  Using the above approximations, the two-photon quantum state is
\begin{equation}
\ket{\psi}=\sum_{\sigma_{s},\sigma_{i}}C_{\sigma_{s},\sigma_{i}}\int\hspace{-2mm}\int\limits_{D}\hspace{-1mm} d\rmvect{q}_{s}
d\rmvect{q}_{i}\ \Phi(\rmvect{q}_{s},\rmvect{q}_{i})\ket{\rmvect{q}_{s},\sigma_{s}}_{s}
\ket{\rmvect{q}_{i},\sigma_{i}}_{i}.
\label{eq:state}
\end{equation}
The kets $\ket{\rmvect{q},\sigma}$ represent single photons in plane wave modes labelled by transverse wavevector $\rmvect{q}$ and polarization $\sigma$.  The subscripts $s$ and $i$ label the down-converted signal and idler modes, respectively.  The coefficients $C_{\sigma_{s},\sigma_{i}}$ depend upon the polarization state of the photon pair.
The function $\Phi(\rmvect{q}_{s},\rmvect{q}_{i})$ is given by\cite{walborn03a}
\begin{equation}
\Phi(\rmvect{q}_{s},\rmvect{q}_{i}) =\frac{1}{\pi}\sqrt{\frac{2L}{K}}\
v(\rmvect{q}_{s}+\rmvect{q}_{i})\
\sinc\left(\frac{L|\rmvect{q}_{s}-\rmvect{q}_{i}|^{2}}{4K} \right),
\label{eq:state2}
\end{equation}
where $v(\rmvect{q})$ is the angular spectrum of the pump beam, which has been transferred to the two-photon quantum state\cite{monken98a}.  $L$ is the length of the nonlinear crystal and $K$ is the magnitude of the pump field wave vector.  For sufficiently thin crystals ($L \ll \, 2Z_{0}$, where $Z_{0}$ is the Rayleigh range of the pump beam), the sinc function is approximately constant.
Depending upon the coefficients $C_{\sigma_{s},\sigma_{i}}$ and the function $\Phi(\rmvect{q}_{s},\rmvect{q}_{i})$,  it is possible that the two-photon state (\ref{eq:state}) is entangled in polarization and/or transverse momentum.  Any entanglement between these two degrees of freedom has been ignored\footnote{There may exist entanglement between polarization and momentum degrees of freedom due to walk off effects in the birrefringent nonlinear crystal.  These effects can be made negligible by the use of additional compensating crystals\cite{kwiat95} as well as interference filters and narrow detection apertures.}.  The function $\Phi(\rmvect{q}_{s},\rmvect{q}_{i})$ is in general a non-separable function of $\rmvect{q}_{s}$ and $\rmvect{q}_{i}$.  As a consequence, the two-photon state must be treated as a single entity.  Emphasizing this fact, the two-photon state is frequently referred to as the \emph{biphoton}.    
\par
In general, two-photon experiments utilize coincidence detections, in which each of two detectors registers the presence of one photon nearly simultaneously.  The probability to detect two photons in coincidence by point-like detectors placed at positions $\brm{r}_1$ and $\brm{r}_2$ is 
\begin{equation}
\mathcal{P}(\brm{r}_1,\brm{r}_2) = |\bsy{\Psi}(\brm{r}_1,\brm{r}_2)|^2,
\label{eq:P}
\end{equation}
where the coincidence detection amplitude, which in the monochromatic approximation serves as the two-photon wave function\cite{mandel95}, is
\begin{equation}
\bsy{\Psi}(\brm{r}_1,\brm{r}_2) =
\bra{\mathrm{vac}}\oper{E}_{2}^{(+)}(\brm{r}_2)\oper{E}_{1}^{(+)}(\brm{r}_1)\ket{\psi},
\label{eq:wf}
\end{equation}
where $\oper{E}_{j}^{(+)}(\brm{r})$ is the field operator and $\brm{r}=(x,y,z)$.  In the paraxial approximation, 
\begin{equation}
\oper{E}_{j}^{(+)}(\brm{r}) = e^{ikz} \sum\limits_{\sigma}\int d\brm{q}
\,\oper{a}_{j}(\brm{q},\sigma)\vect{\epsilon}_{\sigma}e^{i(\brm{q} \cdot
\vect{\rho}-\frac{q^{2}}{2k}z)}.
\label{eq:field}
\end{equation}
The operator $\oper{a}_{j}(\brm{q},\sigma)$ annihilates a photon in mode $j$ with transverse wave vector $\brm{q}$ and polarization $\sigma$.  The vector $\vect{\rho}$ is the transverse component of the position vector $\brm{r}$ and $k$ is the magnitude of the wave vector of the down-converted field.
\par
In general, the coincidence count rate is obtained by integrating the detection probability (\ref{eq:P}) over the area of each detector.  However, in the experiments presented in the following sections, we will be able to analyze the theoretical predictions and experimental results by simply examining $\bsy{\Psi}(\brm{r}_1,\brm{r}_2)$ or $\mathcal{P}(\brm{r}_1,\brm{r}_2)$.    
\section{Multimode Hong-Ou-Mandel Interference}
\label{sec:mmhom}
\begin{figure}[th]
\centerline{\psfig{file=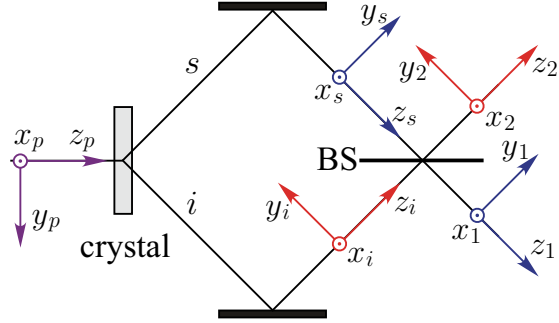,width=3in}}
\vspace*{8pt}
\caption{ \label{fig:mmhom} Illustration of multimode HOM interferometer.}
\end{figure}
An illustration of the multimode HOM interferometer is shown in Fig.
\ref{fig:mmhom}.  Let us assume that the optical path
lengths of photons $s$ and $i$ are equal such that indistinguishable photons suffer fourth-order interference\cite{hom87}.  To
calculate the overall coincidence detection amplitude, it is customary to write
the annihilation operators in modes $1$ and $2$ at the output
ports of the beam splitter in terms of the operators in the input
modes $s$ and $i$:
\begin{align}
\oper{a}_{1}(\brm{q},\sigma)& = t\oper{a}_{s}(q_{x},q_{y},\sigma) + i r
\oper{a}_{i}(q_{x},-q_{y},\sigma) \label{eq:aa} \\
\oper{a}_{2}(q,\sigma)& = t\oper{a}_{i}(q_{x},q_{y},\sigma) + i r
\oper{a}_{s}(q_{x},-q_{y},\sigma),
\label{eq:ab}
\end{align}
where $t$ and $r$ are the transmission and reflection coefficients of the beam splitter.  For simplicity, it has been assumed that the beam splitter is symmetric, though an asymmetric beam splitter will work equally as well.  A key difference between multimode interference and the monomode situation is the sign change in the $y$-components of the reflected wave vectors in Eqs. (\ref{eq:aa}) and (\ref{eq:ab}).  The negative sign is due to the mirror reflection at the beam splitter, as illustrated by the coordinate systems in Fig. \ref{fig:mmhom}.
\par
\begin{figure}[th]
\centerline{\psfig{file=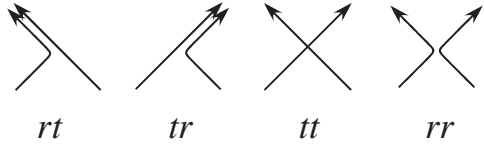,width=2.5in}}
\vspace*{8pt}
\caption{ \label{fig:trans-refl} The four possibilities of transmission and reflection of the down-converted photon pair at the beam splitter.}
\end{figure}

The total detection amplitude contains four terms, corresponding to the possible transmission and reflection of each photon, as illustrated in Fig. \ref{fig:trans-refl}:
\begin{equation}
\label{eq:refl}
\bsy{\Psi} = \bsy{\Psi}_{tr}(\brm{r}_{1},\brm{r}^{\prime}_{1}) +
\bsy{\Psi}_{rt}(\brm{r}_{2},\brm{r}^{\prime}_{2}) +
\bsy{\Psi}_{tt}(\brm{r}_{1},\brm{r}_{2}) +
\bsy{\Psi}_{rr}(\brm{r}_{1},\brm{r}_{2}).
\end{equation}
Here the notation $tr$ stands for transmission of the $s$ photon and reflection of the $i$ photon, etc.  The amplitudes $\bsy{\Psi}_{tr}$ and $\bsy{\Psi}_{rt}$ correspond to two photons in outputs $1$ and $2$ respectively, while $\bsy{\Psi}_{tt}$ and $\bsy{\Psi}_{rr}$ correspond to one photon in each output.  The four components of $\bsy{\Psi}$ are written in
two different coordinate systems, $\brm{r}_{1}=(x_{1},y_{1},z_{1})$ and
$\brm{r}_{2}=(x_{2},y_{2},z_{2})$, since it is necessary to work in the paraxial approximation
around two different axes $z_{1}$ and $z_{2}$. To simplify things, it assumed that $t=r$. Using Eqs. (\ref{eq:state}), (\ref{eq:wf}), (\ref{eq:field}), (\ref{eq:aa}) and (\ref{eq:ab}),  it is straightforward to show that the four components of the detection amplitude are given by\cite{walborn03a}
\begin{align}
\bsy{\Psi}_{tr}(\brm{r}_{1},\brm{r}^{\prime}_{1})=&
i \exp{\left\{\frac{iK}{2Z}\left[(x_{1}-x^{\prime}_{1})^{2}+(y_{1}+
y^{\prime}_{1})^{2}\right]\right\}}
\nonumber \times \label{eq:psitr}\\
&\left[\mathcal{W}\left(\frac{x_{1}+x^{\prime}_{1}}{2},\frac{-y_{1}+y^{\prime}_{1}}{2},Z\right)
  \bsy{\Pi}(\bsy{\sigma}_{1},\bsy{\sigma}^{\prime}_{1})+\right.\nonumber \\
&\left.
\mathcal{W}\left(\frac{x_{1}+x^{\prime}_{1}}{2},\frac{y_{1}-y^{\prime}_{1}}{2},Z\right)
  \bsy{\Pi}(\bsy{\sigma}^{\prime}_{1},\bsy{\sigma}_{1})\right],
\end{align}
\begin{align}
\bsy{\Psi}_{rt}(\brm{r}_{2},\brm{r}^{\prime}_{2})=&
i \exp{\left\{\frac{iK}{2Z}\left[(x_{2}-x^{\prime}_{2})^{2}+
(y_{2}+y^{\prime}_{2})^{2}\right]\right\}}
\nonumber \times \label{eq:psirt}\\
& \left[
\mathcal{W}\left(\frac{x_{2}+x^{\prime}_{2}}{2},\frac{-y_{2}+y^{\prime}_{2}}{2},Z\right)\
\bsy{\Pi}(\bsy{\sigma}_{2},\bsy{\sigma}^{\prime}_{2})+\right.\nonumber \\
&\left.
\mathcal{W}\left(\frac{x_{2}+x^{\prime}_{2}}{2},\frac{y_{2}-y^{\prime}_{2}}{2},Z\right)\
\bsy{\Pi}(\bsy{\sigma}^{\prime}_{2},\bsy{\sigma}_{2})\right],
\end{align}
\begin{align}
\bsy{\Psi}_{tt}(\brm{r}_{1},\brm{r}_{2})=&
\exp{\left\{\frac{iK}{2Z}\left[(x_{1}-x_{2})^{2}+(y_{1}-y_{2})^{2}\right]\right\}}
\nonumber \times\label{eq:psitt}\\
&\mathcal{W}\left(\frac{x_{1}+x_{2}}{2},\frac{y_{1}+y_{2}}{2},Z\right)\
\bsy{\Pi}(\bsy{\sigma}_{1},\bsy{\sigma}_{2}),
\end{align}
and
\begin{align}
\bsy{\Psi}_{rr}(\brm{r}_{1},\brm{r}_{2})=&
- \exp{\left\{\frac{iK}{2Z}\left[(x_{1}-x_{2})^{2}+(y_{1}-y_{2})^{2}\right]\right\}}
\nonumber \times\label{eq:psirr}\\
&\mathcal{W}\left(\frac{x_{1}+x_{2}}{2},\frac{-y_{1}-y_{2}}{2},Z\right)\
\bsy{\Pi}(\bsy{\sigma}_{2},\bsy{\sigma}_{1}).
\end{align}
For simplicity, experimental conditions have been chosen such that each detector is placed at a distance $Z$ from the origin (crystal face) such that $Z = z_{1}=z_{2}$.  The vector $\bsy{\Pi}(\bsy{\sigma}_{1},\bsy{\sigma}_{2})$
is the four-dimensional polarization vector of the input photon pair.  For example, the singlet state is given by $\bsy{\Pi}(\bsy{\sigma}_{1},\bsy{\sigma}_{2}) = 1/\sqrt{2}(\mathbf{h}_{1}\mathbf{v}_{2}-\mathbf{v}_{1}\mathbf{h}_{2})$, where $\mathbf{h}_{j}$ and $\mathbf{v}_{j}$ are orthogonal two-dimensional polarization vectors (for modes $j=1,2$).  The function $\mathcal{W}(x,y,Z)$ is the field profile of the pump beam which has been transferred to the detection amplitude\cite{monken98a}.

For the time being, let us consider only coincidence detections corresponding to one photon in each output of the HOM interferometer.  The detection amplitude for this situation is given by
\begin{equation}
\bsy{\Psi}_{12} = \bsy{\Psi}_{tt}(\brm{r}_{1},\brm{r}_{2}) +
\bsy{\Psi}_{rr}(\brm{r}_{1},\brm{r}_{2}).
\label{eq:psicc}
\end{equation}
Examination of Eqs. (\ref{eq:psitt}) and (\ref{eq:psirr}) shows that the coincidence detection amplitude $\bsy{\Psi}_{12}$ depends upon the parity of the pump field profile $\mathcal{W}$ in the $y$-direction as well as the symmetry of the polarization vector $\bsy{\Pi}$.  For example, if $\mathcal{W}$ is an even function of $y$ and $\bsy{\Pi}$ is symmetric, then $\bsy{\Psi}_{tt}(\brm{r}_{1},\brm{r}_{2})=-\bsy{\Psi}_{rr}(\brm{r}_{1},\brm{r}_{2})$ and the two-photon interference is destructive: $\bsy{\Psi}_{12}=0$.  As expected, in this case $\bsy{\Psi}_{tr}$ and $\bsy{\Psi}_{rt}$ are nonzero, which means that both photons leave through the same port of the HOM interferometer.  However, which port they exit through is entirely random.  On the other hand, if $\mathcal{W}$ is an even function of $y$ and $\bsy{\Pi}$ is antisymmetric, then $\bsy{\Psi}_{tt}(\brm{r}_{1},\brm{r}_{2})=\bsy{\Psi}_{rr}(\brm{r}_{1},\brm{r}_{2})$ and the coincidence detection amplitude is maximum, which corresponds to constructive interference.  Likewise, under these conditions  $\bsy{\Psi}_{tr} = \bsy{\Psi}_{rt}=0$.
\par  
The key feature of  multimode HOM interference is that the multimode treatment of the fields involved in SPDC allows us to consider a pump beam that is an odd function of $y$.  When $\mathcal{W}$ is an odd function of $y$, the interference behavior of the photon pair is reversed:  a symmetric polarization state gives constructive interference and an antisymmetric polarization state results in destructive interference.    
Table \ref{tab:mmhom} summarizes these interference possibilities, which show that the pump beam profile $\mathcal{W}$ can be manipulated to control the interference of symmetric and antisymmetric polarization states.
\begin{table}[pt]
\label{tab:mmhom}
\caption{Summary of two-photon interference dependence in coincidence detections in output $1$ and $2$.}

{\begin{tabular}{|c|c|c|} 
\hline\hline
pump beam profile & polarization state & two-photon interference \\
in $y$ direction & & \\
 \hline
even & symmetric & destructive: $\bsy{\Psi}_{tt} = - \bsy{\Psi}_{rr}$ \\
even  & antisymmetric & constructive: $\bsy{\Psi}_{tt} =  \bsy{\Psi}_{rr}$ \\
odd & symmetric &constructive: $\bsy{\Psi}_{tt} =  \bsy{\Psi}_{rr}$ \\
odd & antisymmetric & destructive: $\bsy{\Psi}_{tt} = - \bsy{\Psi}_{rr}$\\ \hline\hline
\end{tabular} }
\end{table}
\par
To
control the interference behavior of the down-converted fields, it is necessary to pump the nonlinear crystal with a
beam that has well-defined even or odd parity with respect to the $y$-direction.  A  set of beams with well-defined cartesian parity are the
Hermite-Gaussian (HG) beams, given by\cite{beijersbergen93}
\begin{align}
\mathcal{W}_{mn}(x,y,z) = & C_{mn}
H_{m}(x\sqrt{2}/w)H_{n}(y\sqrt{2}/w) e^{-(x^2+y^2)/w^2}  \nonumber \\
& \times e^{-ik(x^2+y^2)/2R(z)} e^{-i(m+n+1)\theta(z)} \nonumber
\end{align}
where $C_{mn}$ is a constant. The $H_{n}(y)$ are the Hermite polynomials, which
are even or odd functions in the $y$-coordinate when the index $n$ is even
or odd, respectively. $w$ is the beam waist,
$R(z) = (z^2+z_{R}^2)/z$ and
$\theta(z) =\arctan(z/z_{R})$,
where $z_{R}$ is the Rayleigh range.
\par
\begin{figure}[th]
\centerline{\psfig{file=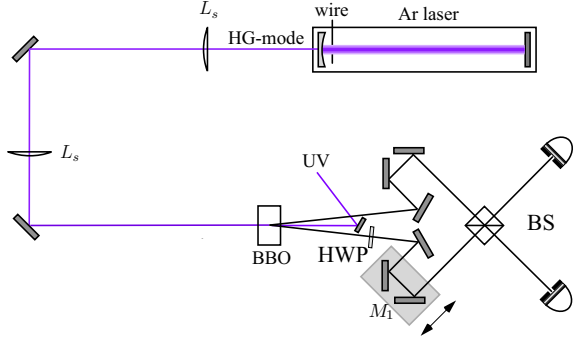,width=3in}}
\vspace*{8pt}
\caption{ \label{fig:mmhom_setup} Experimental setup for multimode Hong-Ou-Mandel interference.  The wire inserted in the laser cavity is used to generate HG modes (see text).}
\end{figure}

The multimode HOM interference experiment\cite{walborn03a} is shown in fig. \ref{fig:mmhom_setup}.  The HG modes were generated by placing a $25\,\mu$m
diameter wire inside
the laser cavity.  The wire breaks the cylindrical symmetry of the laser cavity, which forces the laser to operate in the next stable mode, which is an HG mode with a nodal line at the position of the wire\cite{beijersbergen93}.  Down-converted photons were created by pumping a type-II BBO
($\beta$-BaB$_2$O$_4$) crystal, which generates pairs of orthogonally polarized photons.  Depending on the crystal geometry, these photons may or may not be entangled in polarization\cite{walborn03a,kwiat95}.  The down-converted photons are reflected
through a system of mirrors and incident on a $50-50$ symmetric beam splitter BS ($t =r
\approx \sqrt{1/2}$).   The polarization state is transformed with the half-wave plate HWP.  The symmetric polarization state used was $\ket{\Pi^{S}} =
\ket{h}_{1}\ket{h}_{2}$ and the antisymmetric state was
$\ket{\Pi^{A}} =
\frac{1}{\sqrt{2}}(\ket{h}_{1}\ket{v}_{2}-\ket{v}_{1}\ket{h}_{2})$,
where $h$ and $v$ stands for horizontal and vertical linear
polarization, respectively.  The detectors $D_{1}$ and $D_{2}$
are equipped with interference
filters ($1\,$nm
FWHM centered at $702\,$nm) and $2\,$mm circular apertures, which justifies the use of the monochromatic and paraxial approximations in the basic theory presented above.  
\par
\begin{figure}[th]
\centerline{\psfig{file=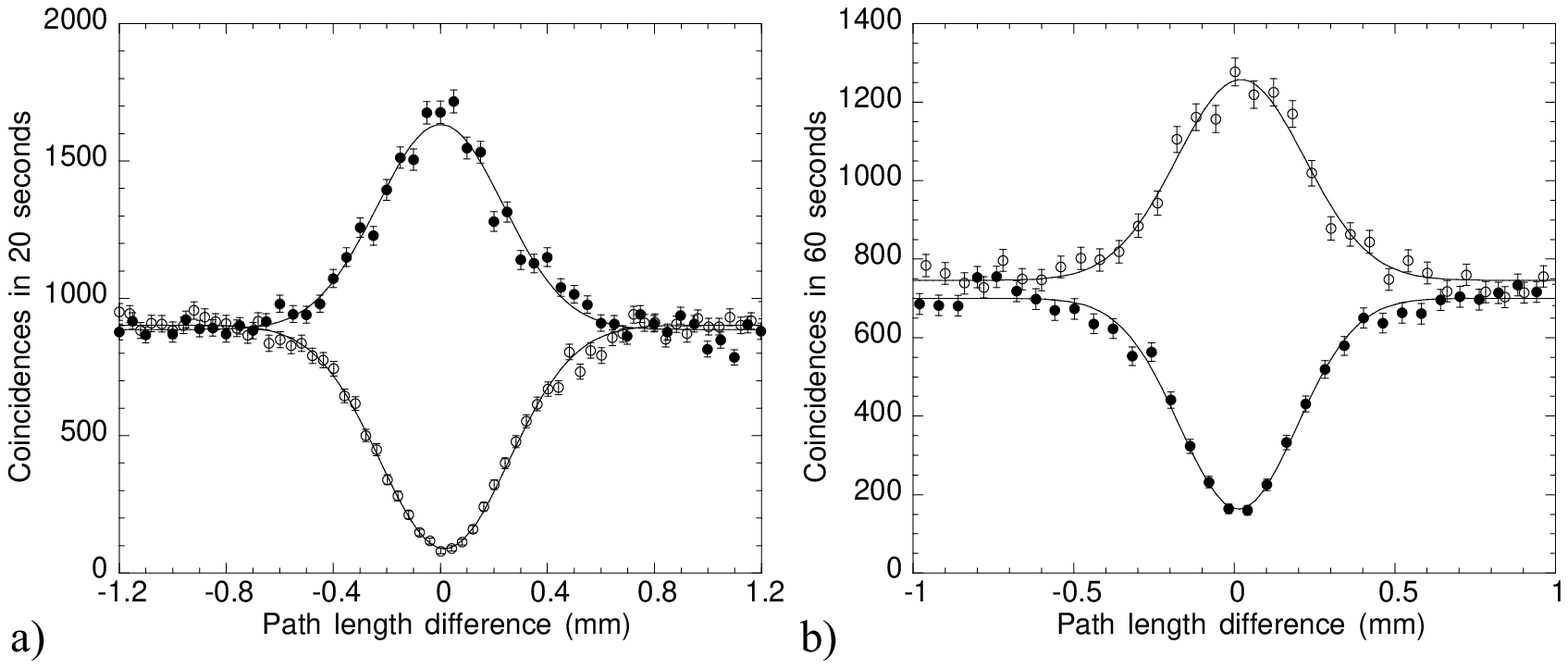,width=5in}}
\vspace*{8pt}
\caption{\label{fig:mmhom_results} Experimental results for multimode Hong-Ou-Mandel interference for a) symmetric polarization state and b) antisymmetric polarization state.  Open circles correspond to the HG$_{10}$ and solid circles correspond to the HG$_{01}$ pump beam.   }
\end{figure}
\begin{figure}[th]
\centerline{\psfig{file=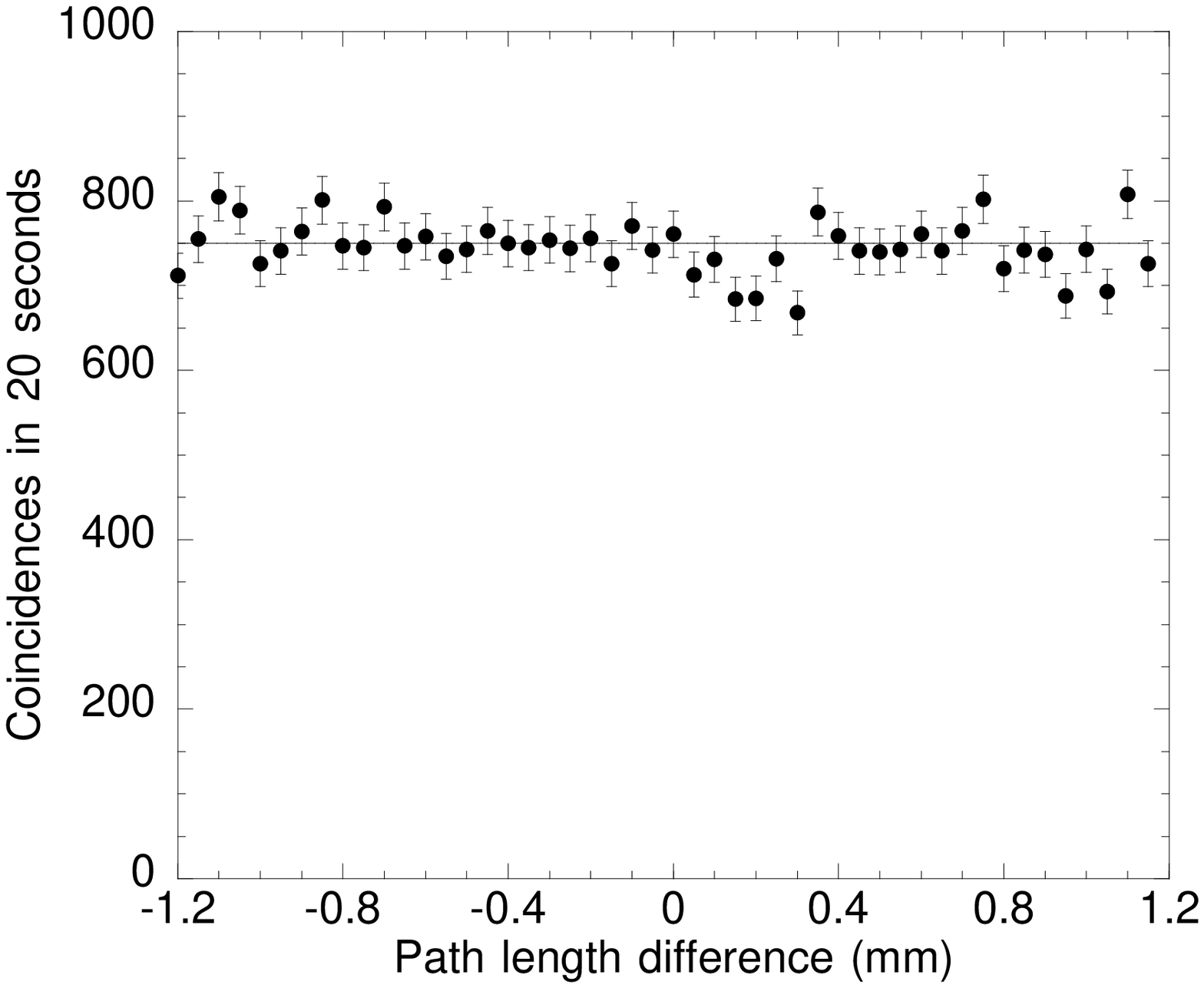,width=2.5in}}
\vspace*{8pt}
\caption{Experimental results for multimode Hong-Ou-Mandel interference for symmetric polarization state and a pump beam with undefined parity.   \label{fig:DHG}}
\end{figure}

Figs. \ref{fig:mmhom_results} and \ref{fig:DHG} show the HOM interference curves measured by detecting coincidences as a function of the path length difference, which was controlled by scanning the mirror assembly $M_{1}$ with a linear stepper motor.  The error bars represent statistical errors due to photon counting\cite{mandel95} and the solid lines are curve fits of the usual HOM interference curve\cite{hom87}.  Fig. \ref{fig:mmhom_results} a) shows results for the symmetric polarization state $\ket{\Pi^{S}}$ for the $\mathcal{W}_{10}$ (open circles) and $\mathcal{W}_{01}$ (solid circles) pump beams.   Fig. \ref{fig:mmhom_results} a) shows results for the antisymmetric polarization state $\ket{\Pi^{A}}$ for the $\mathcal{W}_{10}$ (open circles) and $\mathcal{W}_{01}$ (solid circles) pump beams.  When the path length difference is zero, these results confirm the theoretical predictions summarized in table \ref{tab:mmhom}.   
Creating a pump beam that is an equally weighted superposition of $\mathcal{W}_{10}$ (even) and $\mathcal{W}_{01}$ (odd) modes results in a beam with undefined parity in the $y$ direction.  The beam was created by aligning the wire in the laser cavity at a $45^\circ$ angle.
No HOM interference is observed when the crystal is pumped with such a beam, as is shown in Fig. \ref{fig:DHG} for the symmetric polarization state.  
\section{Biphoton phase measurements}
A possible application of multimode HOM interference is the measurement of the transverse phase structure of the biphoton.  Such a measurement can be used to show that down-converted photon pairs are entangled in orbital angular momentum\cite{walborn04}.  
\par
Suppose that the transverse profile of the pump beam is a Laguerre-Gaussian (LG) beam.  It is well known that the LG beams carry orbital angular momentum in the form of an azimuthal phase dependence $\exp(i l \phi)$, where $\phi$ is the azimuthal angle and $l$ is the azimuthal winding number.  It has been shown that LG beams carry an angular momentum of $l \hbar$ per photon\cite{allen92}.    An important issue is whether SPDC can be used to create photons entangled in orbital angular momentum, and if so, what are the experimental requirements necessary to do so?  These questions have been the focus of much recent work\cite{mair01,arnaut01,barbosa02,franke-arnold02,torres03}.  Photons pairs entangled in orbital angular momentum might be very useful in quantum communication and quantum cryptography\cite{gisin02}, since they allow for the encoding of higher order alphabets.   
\par
Consider for a moment the simple case of a nonlinear crystal pumped with a LG beam.  In this case, the two-photon wave function, in the absence of a HOM interferometer, is         
\begin{equation}
\Psi(\bsy{\rho}_{s},\bsy{\rho}_{i}) = \mathcal{U}_{p}^{l}\left(\frac{\bsy{\rho}_{s}+\bsy{\rho}_{i}}{\sqrt{2}}\right),
\label{eq:psif2}
\end{equation}
where we have used the paraxial, monochromatic and thin crystal approximations.  Here $\mathcal{U}_{p}^{l}$ is the field profile of the LG beam with radial index $p$ and azimuthal index $l$.  The polarization state has been assumed to be symmetric, which allows the biphoton wave function to be treated as a scalar.  The LG modes are given by       
\begin{align}
\mathcal{U}_{p}^{l}(\rho,\phi,z)=&D_{lp}
\left(\frac{\sqrt{2}\rho}{w(z)}\right)^{|l|}L_{p}^{|l|}\left(\frac{2\rho^2}{w(z)^2}\right)
\exp\left(-\frac{\rho^2}{w(z)}\right)\times \nonumber \\
&\exp\left(-i\left(\frac{k\rho^2}{2R}
-(2p+|l|+1)\theta(z)\right)-il\phi\right)
\label{eq:LG}
\end{align}
where $(\rho,\phi,z)$ are the usual cylindrical coordinates, $D_{lp}$ is a constant and $L_{p}^{l}$ are the Laguerre polynomials.  The order of the LG beam is  $\mathcal{N} = |l|+2p$.

\par
The LG modes form a complete basis, so the two photon state can be expanded in terms of LG modes of the signal and idler fields:
\begin{equation}
\Psi(\bsy{\rho}_{s},\bsy{\rho}_{i})=\sum_{l_{s},p_{s}}\sum_{l_{i},p_{i}}C_{p_{s}p_{i}}^{l_{s}l_{i}}\mathcal{U}_{p_{s}}^{l_{s}}(\bsy{\rho}_{s})\,\mathcal{U}_{p_{i}}^{l_{i}}(\bsy{\rho}_{i}).
\label{eq:LGmode}
\end{equation}
It has been shown that\cite{walborn04} 
\begin{equation}
C_{p_{s}p_{i}}^{l_{s}l_{i}}\propto \delta_{l_{s}+l_{i},l}\int\hspace{-1mm} q\,dq\   v_{p}^{l}(\sqrt{2}\,q)\,v_{p_{s}}^{*l_{s}}(q)\,v_{p_{i}}^{*l_{i}}(q),
\label{eq:cfour2}
\end{equation} 
where $v_{p}^{l}(q)$ is the Fourier transform of the LG mode given by Eq. (\ref{eq:LG}).
The conservation of orbital angular momentum in SPDC is provided by the delta function in Eq. (\ref{eq:cfour2}), which guarantees that $l=l_{s}+l_{i}$, provided that Eq. (\ref{eq:psif2}) indeed describes the biphoton wave function.    
\par
It has been shown that Eq. (\ref{eq:psif2}) accurately describes the two-photon state through a simple experiment.  Direct coincidence detection provides information about the modulus of $\Psi(\bsy{\rho}_{s},\bsy{\rho}_{i})$, while multimode HOM interference provides information about the phase structure.  
Suppose that the HOM interferometer described in section \ref{sec:mmhom} is pumped by a LG beam.  The corresponding coincidence detection amplitude is\cite{walborn04}
\begin{equation}
\Psi_{c}(\bsy{\rho}_{1},\bsy{\rho}_{2})=\Psi_{c}(R,\theta)\propto\, u_{p}^{l}(R)\sin l\theta,
\label{eq:ampli}
\end{equation}
where $R=\frac{1}{\sqrt{2}}|\bsy{\rho}_{1}+\bsy{\rho}_{2}|$ and $\theta$  is defined by the relations
\begin{align}
\sin{\theta}=\frac{\rho_{1}\sin{\phi_{1}}+\rho_{2}\sin{\phi_{2}}}{R}\\
\cos{\theta}=\frac{\rho_{1}\cos{\phi_{1}}+\rho_{2}\cos{\phi_{2}}}{R}.
\end{align}
Here $u_{p}^{l}$ contains all of the radial dependence of $\mathcal{U}_{p}^{l}$.   
The coincidence detection probability, which is proportional to $|\Psi_{c}(R,\theta)|^{2}$, is
\begin{equation}
P_{12}(\bsy{\rho}_{1},\bsy{\rho}_{2})\propto |u_{p}^{l}(R)|^{2}\sin^{2} l\theta.
\label{eq:prob1}
\end{equation}
\begin{figure}[th]
\centerline{\includegraphics[width=3in]{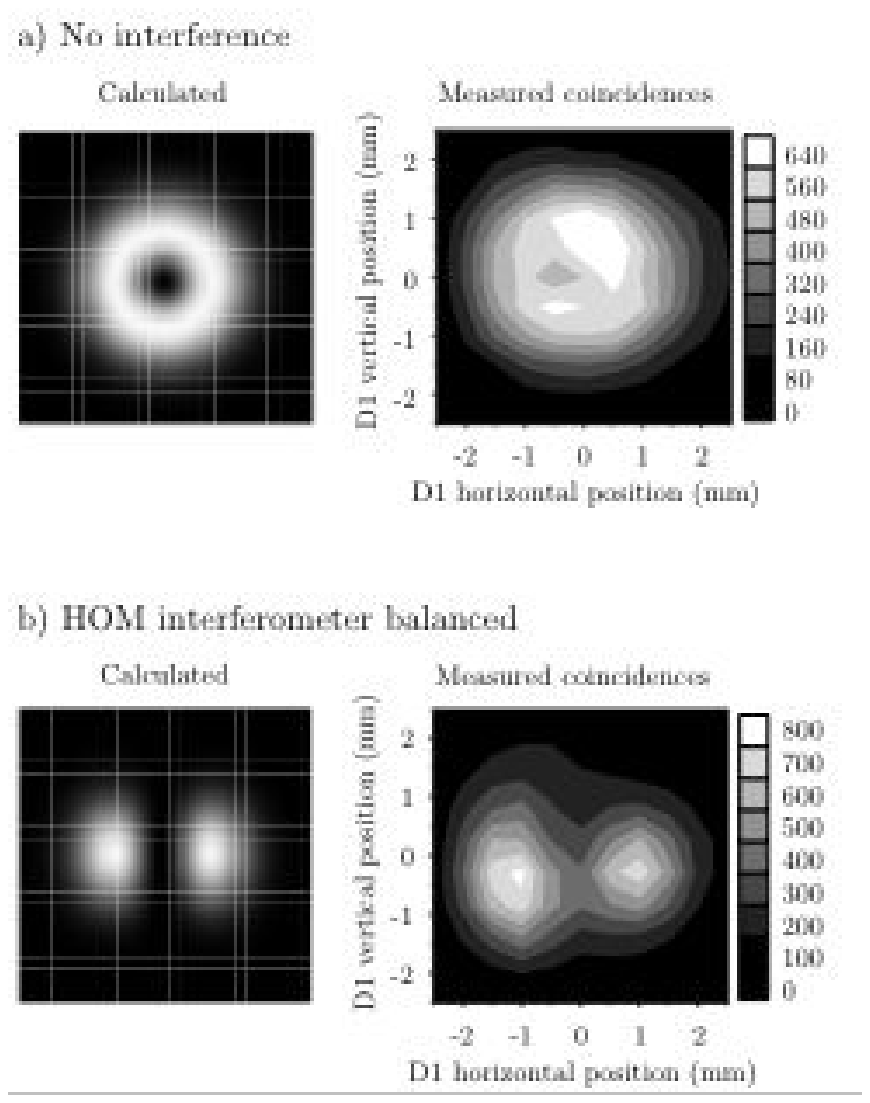}}
\vspace*{8pt}
\caption{Coincidence profiles predicted (left) and measured (right) when the crystal is pumped by a LG$_{0}^{1}$ beam. a) No-interference regime (Hong-Ou-Mandel interferometer unbalanced). b) Fourth-order interference regime (interferometer balanced).  \label{fig:lghom2}}
\end{figure}
\begin{figure}[th]
\centerline{\includegraphics[width=3in]{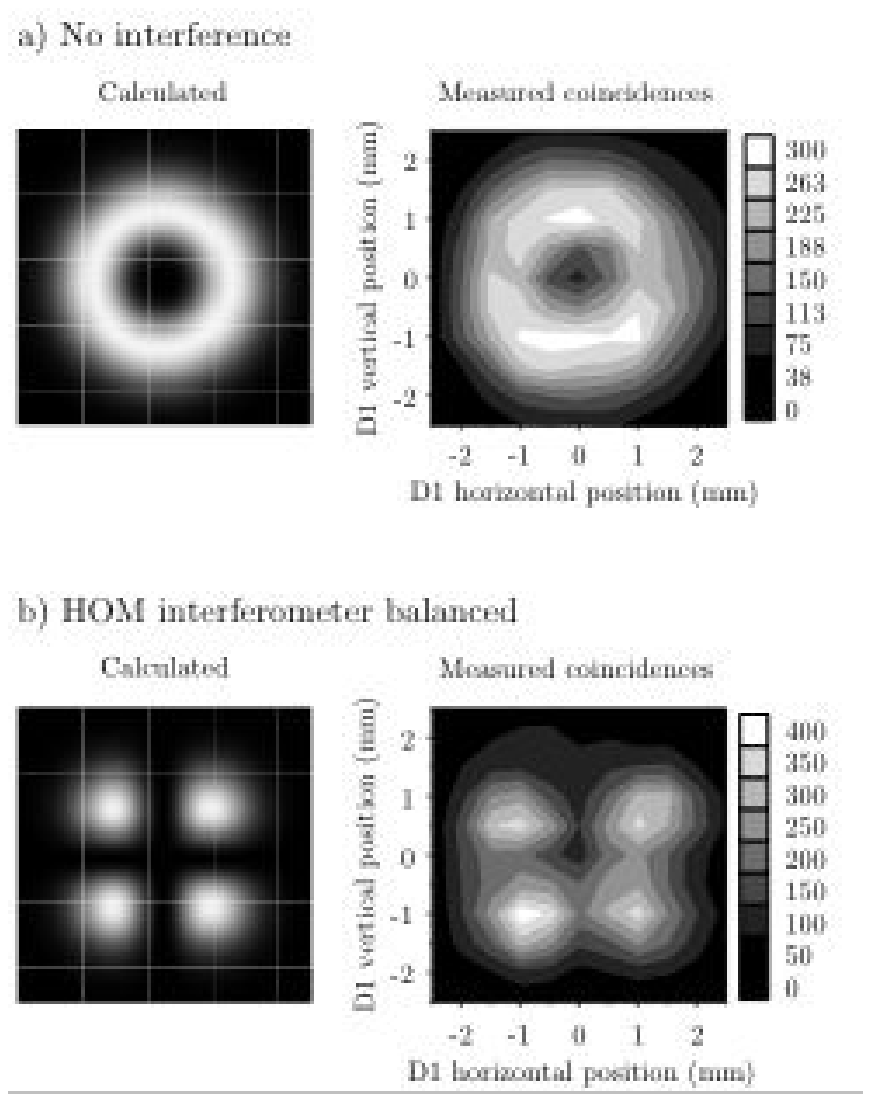}}
\vspace*{8pt}
\caption{Coincidence profiles predicted (left) and measured (right) when the crystal is pumped by a LG$_{0}^{2}$ beam. a) No-interference regime (Hong-Ou-Mandel interferometer unbalanced). b) Fourth-order interference regime (interferometer balanced)..  \label{fig:lghom1}}
\end{figure}
\par
Figs. \ref{fig:lghom2} and \ref{fig:lghom1} show experimental results for $\mathcal{U}_{0}^{1}$ and $\mathcal{U}_{0}^{2}$ pump beams, respectively.  
Using a mode converter consisting of a pair of cylindrical lenses\cite{beijersbergen93}, first and second order HG beams (created by inserting a wire into the laser cavity) were transformed into LG beams of the same order.  The LG beams were then used to pump the HOM interferometer.  Coincidence counts were registered by setting one detector (1 mm diameter circular aperture) fixed while the other detector (0.5 mm diameter circular aperture) mapped a 2-dimensional grid in the transverse plane.   In part a) of the figures, the HOM interferometer was ``unbalanced", meaning that the path length difference was greater than the coherence length of the down-converted photons.  Under these conditions, there is no interference, and the coincidence profile reproduces the field profile of the pump beam.  Parts b) show the interference when the HOM interferometer is balanced (equal path lengths).  Comparing the experimental results with a computer simulation of Eq. (\ref{eq:prob1}) shows good agreement between the experimental and theoretical coincidence distributions.  The above experimental results show that  Eq. (\ref{eq:psif2}) accurately describes the two-photon state, which in turn shows that OAM is conserved in the SPDC process.  
\par
\begin{figure}[th]
\centerline{\psfig{file=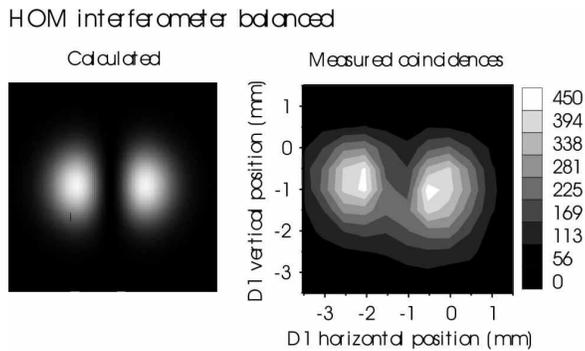,width=3in}}
\vspace*{8pt}
\caption{Coincidence profile predicted (left) and measured (right) when the crystal is pumped by a LG$_{0}^{1}$ beam, in the fourth-order interference regime (Hong-Ou-Mandel interferometer balanced). Detector $D_{2}$ was displaced by $\Delta x = \Delta y = 1$ mm.  Note that the coincidence pattern is shifted by $-1$\, mm in the $x$ and $y$ directions.\label{fig:lghom3}}
\end{figure}

 Through a simple argument, it can be shown that the down-converted fields are indeed entangled in OAM.   
From Eq. (\ref{eq:psif2}) it is evident that $\Psi(\bsy{\rho}_{s}+\bsy{\Delta},\bsy{\rho}_{i}-\bsy{\Delta})=\Psi(\bsy{\rho}_{s},\bsy{\rho}_{i})$.  Fig. \ref{fig:lghom3} shows experimental results when the fixed detector was displaced  $\Delta_{x}=\Delta_{y}=1$\,mm.  The interference pattern is equivalent to that of Fig. \ref{fig:lghom2} b) but shifted by $-1$\,mm in the $x$ and $y$ directions, as predicted by Eq. (\ref{eq:psif2}).       
Due to the singular phase structure of $\mathcal{U}_{p}^{l}$, for  $l \neq 0$ there exist transverse spatial positions $\bsy{\rho}_{s0}$ and $\bsy{\rho}_{i0}$ such that \begin{equation}
\Psi(\bsy{\rho}_{s0}+\bsy{\Delta},\bsy{\rho}_{i0}-\bsy{\Delta})=\Psi(\bsy{\rho}_{s0},\bsy{\rho}_{i0}) = 0,
\end{equation}
and the coincidence detection probability $\mathcal{P}(\bsy{\rho}_{s},\bsy{\rho}_{i}) = |\Psi(\bsy{\rho}_{s},\bsy{\rho}_{i})|^{2}$ is 
 \begin{equation}
 \mathcal{P}(\bsy{\rho}_{s0}+\bsy{\Delta},\bsy{\rho}_{i0}-\bsy{\Delta})=\mathcal{P}(\bsy{\rho}_{s0},\bsy{\rho}_{i0}) = 0.  
 \label{eq:new2}
 \end{equation} 
 \par
Besides entanglement, which is a purely quantum correlation, the only other way for OAM to be conserved in SPDC would be through a classical correlation\cite{peres95} of the signal and idler fields.  Let us suppose that the down-converted fields exhibit such a classical correlation.    The detection probability $\mathcal{P}_{\mathrm{cc}}$ for this type of correlation can be written as
\begin{equation}
\mathcal{P}_{\mathrm{cc}}(\bsy{\rho}_{s},\bsy{\rho}_{i}) = \sum\limits_{l_{i}=-\infty}^{\infty}P_{l_{i}}|F_{l-l_{i}}(\bsy{\rho}_{s})|^{2}|G_{l_{i}}(\bsy{\rho}_{i})|^{2},
\label{eq:new3}
\end{equation}
where $F_{l_{s}}(\bsy{\rho}_{s})$ and $G_{l_{i}}(\bsy{\rho}_{i})$ represent down-converted fields with average orbital angular momentum 
$l_{s} \hbar$ and $l_{i} \hbar$ per photon.  Since the correlation is classical,  the coefficients $P_{l_{i}}$ satisfy $\sum_{l_{i}=-\infty}^{\infty}P_{l_{i}} = 1$ and $P_{l_{i}} \geq 0$.  For Eq. (\ref{eq:new3}) to accurately describe the two-photon state, it must satisfy the equivalent of Eq. (\ref{eq:new2}):
 \begin{equation}
 \mathcal{P}_{\mathrm{cc}}(\bsy{\rho}_{s0}+\bsy{\Delta},\bsy{\rho}_{i0}-\bsy{\Delta})  =\mathcal{P}_{\mathrm{cc}}(\bsy{\rho}_{s0},\bsy{\rho}_{i0})  = 0, \nonumber 
 \end{equation}
 which gives 
 \begin{equation}
\sum\limits_{l_{i}=-\infty}^{\infty}P_{l_{i}}|F_{l-l_{i}}(\bsy{\rho}_{s0}+\bsy{\Delta})|^{2}|G_{l_{i}}(\bsy{\rho}_{i0}-\bsy{\Delta})|^{2} =0. 
 \label{eq:new4}
 \end{equation}  
 Since $P_{l_{i}} \geq 0$, a solution to Eq. (\ref{eq:new4}) exists (for the non-trivial cases where at least one $P_{l_{i}}\neq 0$) only if $|F_{l-l_{i}}(\bsy{\rho}_{s0}+\bsy{\Delta})|^{2} = 0$ or $|G_{l_{i}}(\bsy{\rho}_{i0}-\bsy{\Delta})|^{2} = 0$ for all $\bsy{\Delta}$. This implies that $F_{l-l_{i}}\equiv 0$ or $G_{l_{i}}\equiv 0$.  Thus, a classical
correlation of orbital angular momentum  states cannot
reproduce the two photon wave function (\ref{eq:psif2}), which indicates that the down-converted fields are indeed entangled in OAM.  We note here that the above argument is valid even though the systems under consideration are (in principle) infinite dimensional. 
 \section{Two-photon singlet beams}
\label{sec:singbeams}
In the previous sections we were concerned with coincidence counts in different outputs of the HOM interferometer.  Here we will consider situations in which both photons leave through the same output port.
\par
An interesting application of multimode interference is the creation of a localized two-photon singlet state\cite{nogueira04}.  Such a state can be generated by controlling the HOM interference such that two input photons in the singlet polarization state leave the beam splitter through the same port.
  In this type of beam, both the polarization and spatial components of the two-photon wave function are antisymmetric, which gives rise to some interesting properties.  First, it is known that, in general, the singlet polarization state $\ket{\psi^{-}}$ is invariant to any bilateral unitary operator $\oper{U}_{\mathrm{T}}=\oper{U}\otimes\oper{U}$  \cite{kwiat00,cabello03}.  Suppose that this operator characterizes some type of noise in a quantum channel.  Then it could be possible to use the singlet state (along with another invariant state) to send quantum information in a robust manner.  The assumption that the noise is bilateral is generally justified as long as the photons occupy the same spatio-temporal region such as, in this case, a well-collimated beam. 
  \par
 A second interesting property is that the spatial antisymmetry guarantees that the singlet-beam exhibits spatial antibunching in the transverse plane.  It has been shown that spatial antibunching is a purely quantum property with no classical analog\cite{nogueira01,nogueira02}.  Thus the singlet-beam is inherently non-classical.  
\par
To generate a singlet beam, a pump beam that is an odd function of the $y$ coordinate $\mathcal{W}_{\mathrm{odd}}(x,-y,z) = -\mathcal{W}_{\mathrm{odd}}(x,y,z)$ is used.  When the polarization state is $\ket{\psi^{-}}$,  the probability
amplitude to detect \emph{both} photons in output port 1 is
given by
\begin{align}\label{eq:sing}
\bm{\Psi}(\mathbf{r}_{1},\mathbf{r}_{1}^{\prime}) \propto
 & \, \mathcal{W}_{\mathrm{odd}} \left(\frac{x_{1}+x_{1}^{\prime}}{2},\frac{y_{1}-y_{1}^{\prime}}{2},Z \right) \nonumber \\
& \times  (\brm{h}_{1} \brm{v}_{1}^{\prime}- \brm{v}_{1} \brm{h}_{1}^{\prime})
\end{align}
where $\mathbf{r}_{1} = (x_{1},y_{1},z_{1})$ and
$\mathbf{r}_{1} = (x_{1}^{\prime},y_{1}^{\prime},z_{1}^{\prime})$ are the coordinates of
detectors $D_{1}$ and $D_{1}^{\prime}$, respectively, with $z_{1} = z_{1}^{\prime} = Z$.    Here both detectors are placed in the
same output port of the BS, \textit{i.e.}, $D_{1}$ and $D_{1}^{\prime}$ detect in the same
spatial region. 
\begin{figure}[th]
\centerline{\psfig{file=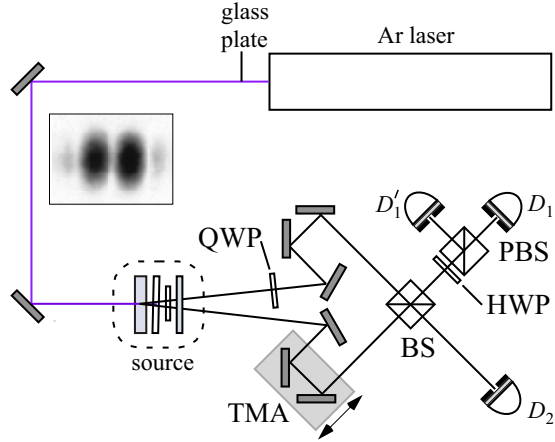,width=3in}}
\vspace*{8pt}
\caption{ \label{fig:sing_setup} Experiment for generation and detection of a two-photon singlet beam.}
\end{figure}
\par
To generate a pump beam that is an odd function of the $y$ coordinate, a thin ($\sim 150\,\mu$m) glass laminate was inserted halfway into the Gaussian profile pump beam and adjusted the angle in order to achieve a  $\pi$ phase difference
between the two halves of the beam.  In the far field the beam profile is approximately equal to $\mathcal{W}_{01}$.
This beam is used to pump a nonlinear crystal
which is
adjusted to generate polarization-entangled photons using the
crossed-cone source\cite{kwiat95}.  The output
state of this source is controlled by adjusting the angle of the compensating crystal
to be the $| \psi^{+} \rangle$ polarization
state. With a quarter-wave plate (QWP) in one of the paths,
the relative phase can be manipulated in order to change from the polarization state
$| \psi^{+} \rangle$ to $| \psi^{-} \rangle$\cite{kwiat95}.  
 The quality of the polarization-entangled state can be tested in the usual way using polarization analyzers\cite{kwiat95}.
 The entangled photons are then sent into the HOM interferometer.  
\par
Coincidences were measured at the two
output ports of a polarizing beam splitter placed in one output of the HOM interferometer, so that detectors $D_{1}$ and $D_{1}^{\prime}$ always detect orthogonal polarizations.
The path length difference was scanned as in the usual HOM interference measurements, however, this time
 coincidences were registered at detectors in the same output port of the BS.   
 \par
\begin{figure}[th]
\centerline{\psfig{file=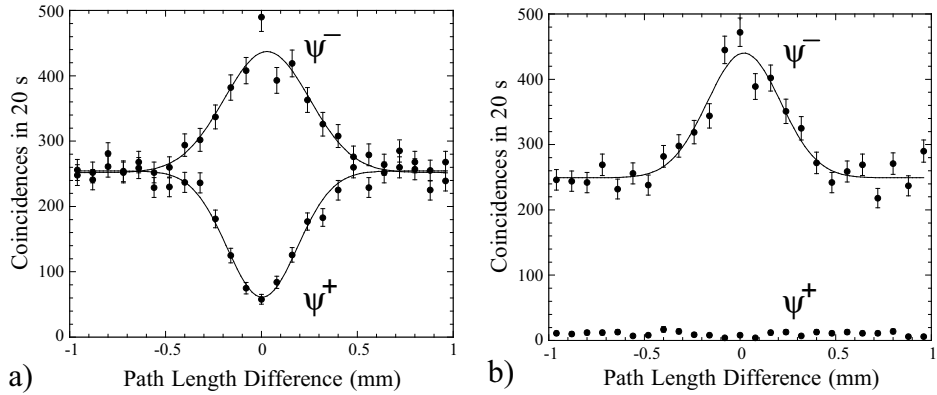,width=5in}}
\vspace*{8pt}
\caption{Experimental results for two-photon singlet and triplet beams for  a)  h/v measurement basis and b) +/- measurement basis.   \label{fig:sing_res1}}
\end{figure}

 For the $\ket{\psi^{-}}$ state,  constructive interference is observed at detectors $D_{1}$ and $D_{1}^{\prime}$ in both the $h/v$ (Fig. \ref{fig:sing_res1}a) and $+/-$ bases, $\pm = 1/\sqrt{2}(h \pm v)$ (Fig. \ref{fig:sing_res1}b).  Observing constructive interference in both detection bases identifies the $\ket{\psi^{-}}$ state, since it is the only antisymmetric  two-photon polarization state and is invariant to bilateral rotation.  
 \par
 It is illustrative to  compare these results with those of the input $\ket{\psi^{+}}=1/\sqrt{2}\left(\ket{h}_{1}\ket{v}_{2} +
\ket{v}_{1} \ket{h}_{2} \right)$ polarization state. If the pump beam is an even function of $y$ and the polarization state is $\ket{\psi^{+}}$, the detection amplitude is
\begin{align}\label{eq:trip}
\bm{\Psi}(\mathbf{r}_{1},\mathbf{r}_{1}^{\prime}) \propto
 & \, \mathcal{W}_{\mathrm{even}} \left(\frac{x_{1}+x_{1}^{\prime}}{2},\frac{y_{1}-y_{1}^{\prime}}{2},Z \right) \nonumber \\
& \times  (\brm{h}_{1} \brm{v}_{1}^{\prime}+ \brm{v}_{1} \brm{h}_{1}^{\prime}).
\end{align}
Detecting in the $h/v$ basis, we observe an interference ``dip" (Fig. \ref{fig:sing_res1}a). However, in the
$+/-$ basis (Fig. \ref{fig:sing_res1} b), we observe no coincidences, since in this basis
 the $| \psi^{+} \rangle$ state is proportional to $(\ket{+}_{1}\ket{+}_{1}^{\prime} - \ket{-}_{1}\ket{-}_{1}^{\prime})$.
\par
\section{Optical Bell-State Measurement}
\label{sec:obsa}
The four entangled Bell-states:
\begin{align}
& \ket{\psi^{\pm}}=\frac{1}{\sqrt{2}}\left(\ket{h}_{1}\ket{v}_{2} \pm
\ket{v}_{1} \ket{h}_{2} \right), \nonumber \\
& \ket{\phi^{\pm}}=\frac{1}{\sqrt{2}}\left(\ket{h}_{1}\ket{h}_{2} \pm
\ket{v}_{1} \ket{v}_{2} \right), \label{eq:bellstates}
\end{align}
play a crucial role in many quantum information schemes.   Here
$\ket{\psi^{-}}$ is the antisymmetric singlet state and
$\ket{\psi^{+}}$,$\ket{\phi^{\pm}}$ are the symmetric triplet
states.  Furthermore, many quantum information schemes, such as
dense coding\cite{mattle96,bennett92}, quantum teleportation\cite{bennett93} and entanglement swapping\cite{bennett93} require a Bell-state
measurement (BSM), that is, projecting onto the basis defined by
the states (\ref{eq:bellstates}). Generally,  optical BSM's
\cite{mattle96,braunstein95} of polarization-entangled photons
rely on HOM-type interference at a 50-50
beam splitter\cite{hom87}.
\begin{figure}[th]
\centerline{\psfig{file=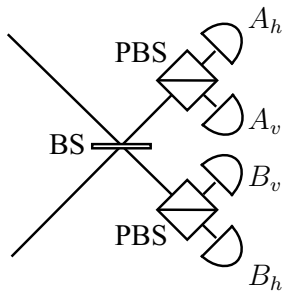,width=4cm}} \vspace*{8pt}
\caption{Illustration of an optical Bell-state
analyzer.\label{fig:BSA}}
\end{figure}
A typical optical Bell-state analyzer (BSA) is shown in Fig.
\ref{fig:BSA}.   For the moment, let us consider only monomode fields, as given in Eq. (\ref{eq:bellstates}).  A 50-50 beam splitter is used to separate
$\ket{\psi^{-}}$ from $\ket{\psi^{+}}$, $\ket{\phi^{\pm}}$. Bosonic symmetry requires
that photons in the $\ket{\psi^{-}}$ state end up in different
outputs while photons in $\ket{\psi^{+}}$, $\ket{\phi^{\pm}}$ end
up in the same output\cite{braunstein95}.
 With the polarizing beam splitters (PBS) separating $h$ and $v$
polarizations, coincidences at $A_{h}B_{v}$ or $A_{v}B_{h}$
identify the $\ket{\psi^{-}}$ polarization state.  The PBS also
separate the $\ket{\psi^{+}}$ state from the $\ket{\phi^{\pm}}$
states: coincidences at $A_{h}A_{v}$ or $B_{h}B_{v}$ are the
signature of the $\ket{\psi^{+}}$ state.  For $\ket{\phi^{\pm}}$,
we have two-photon detections of the form $A_{h}A_{h}$,
$A_{v}A_{v}$, $B_{h}B_{h}$ or $B_{v}B_{v}$. This scheme allows for identification of 3 classes of states, however, detection of
$\ket{\psi^{-}}$, $\ket{\psi^{+}}$ and $\ket{\phi^{\pm}}$ requires
detectors capable of distinguishing between one and two photons.
Such detectors are presently available, however they suffer from
low efficiencies and/or high dark counts
\cite{kwiat93,kim99,takeuchi99}.  This problem can be partially solved by
replacing each detector an additional 50-50 beam splitter and two
detectors\cite{mattle96}. This enables one to detect only half of the two-photon
occurrences and increases the complexity of the detection system,
since an eight detector system is necessary.  Recently developed multi-photon detectors based on temporal multiplexing would reduce the number of detectors in this situation, but would not operate with 100\% efficiency\cite{achilles04}.     
  \par
This requirement on the detectors can be avoided if the interference behavior is inverted:  photons in the triplet (singlet) states go to  different (the same) detectors and can then be further discriminated by the PBS.  As shown above, this can be achieved by generating polarization-entangled photons using an
antisymmetric pump beam, such as the first-order Hermite-Gaussian
beam $\mathcal{W}_{01}$.
 Then, pumping with $\mathcal{W}_{01}$,
$\ket{\psi^{-}}$ results in
two photons in either output port.  Since the two photons are
orthogonally polarized, coincidences at detectors $A_{h}A_{v}$ or
$B_{h}B_{v}$ identify the $\ket{\psi^{-}}$ state.   The states $\ket{\psi^{+}}$ and
$\ket{\phi^{\pm}}$ give one photon in each output port.  Since the
photon pairs of $\ket{\psi^{+}}$ are orthogonally polarized, $\ket{\psi^{+}}$
gives coincidence counts  at detectors  $A_{h}B_{v}$ or
$B_{h}A_{v}$.   $\ket{\phi^{\pm}}$ results in coincidence counts at
$A_{h}B_{h}$ or $A_{v}B_{v}$.       All  detector combinations identifying the three cases $\ket{\psi^{+}}$, $\ket{\psi^{-}}$ and $\ket{\phi^{\pm}}$ correspond to coincidences at different detectors.  Three classes of states can be identified with only 4 detectors.  
\par
\begin{figure}[th]
\centerline{\psfig{file=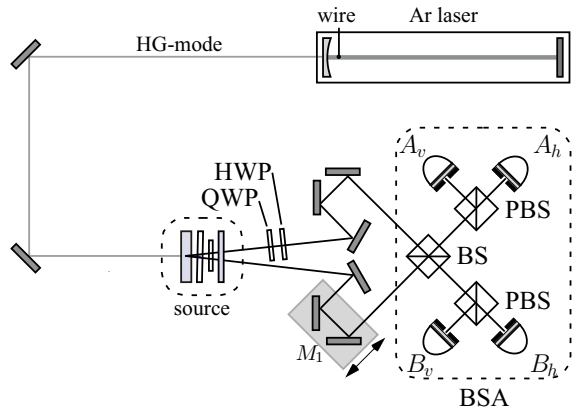,width=3in}}
\vspace*{8pt}
\caption{Experimental setup of multimode optical Bell-state analysis.  The setup is the same as that of Fig. \ref{fig:mmhom_setup} with a polarizaton Bell-state source\cite{kwiat95} and the inclusion of the Bell-state analyzer shown in Fig. \ref{fig:BSA}. \label{fig:BSMsetup}}
\end{figure}
\begin{figure}[th]
\centerline{\psfig{file=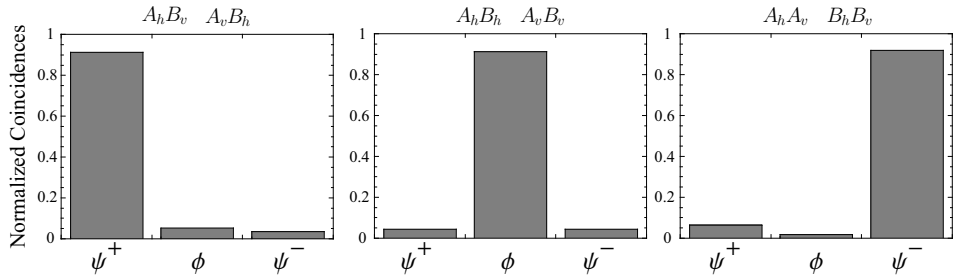,width=5in}}
\vspace*{8pt}
\caption{Experimental results for multimode optical Bell-state analysis.\label{fig:BSM}}
\end{figure}

Fig. \ref{fig:BSMsetup} shows the experimental setup used to discriminate three classes of Bell-states in coincidence detections\cite{walborn03b}.  Results of the experiment, presented in Fig. \ref{fig:BSM}, show that the three classes of states were identified with different coincidence counts at different detector combinations.  The analyzer functioned with an efficiency of about $90\%$, with experimental errors due mostly to alignment of the HOM interferometer.   
\section{Conclusion}
We have reviewed recent experiments centered around multimode Hong-Ou-Mandel interference.  A key ingredient is the multimode treatment of spontaneous parametric down-conversion in the monochromatic, paraxial and thin-crystal approximations, which guarantees that the angular spectrum of the pump laser field is transferred to the two-photon state.  As a consequence, the transverse spatial characteristics of the pump beam can be manipulated in order to control the fourth-order interference behavior of photon pairs in the Hong-Ou-Mandel interferometer. 
\par
Applications of multimode Hong-Ou-Mandel interference to the field of quantum information include measurements of the transverse phase characteristics of the two-photon state, generation of two-photon singlet beams, and simplified Bell-state measurement of down-converted photon pairs.  We expect that multimode interference will play also a role in small-scale quantum logic gates, quantum communication and quantum imaging.  Future work might include multimode interference of more than two photons, as well as photons originating from different sources.    
\section*{Acknowledgments}
We would like to thank R. S. Thebaldi for insightful discussions and M. P. Almeida for a careful reading of this manuscript.  
The authors acknowledge financial support from the Brazilian funding agencies CNPq and CAPES.

\end{document}